\documentclass[12pt,preprint,preprintnumbers,nofootinbib,groupedaddress,superscriptaddress,amsmath,amssymb]{revtex4}
\usepackage{graphicx}
\usepackage{dcolumn}
\usepackage{bm}
\usepackage{amssymb}
\usepackage{amsmath}
\usepackage{epsfig}    
\usepackage{color}
\usepackage{hhline}

\def\be{\begin{equation}}
\def\ee{\end{equation}}
\newcommand{\bea}{\begin{eqnarray}}
\newcommand{\eea}{\end{eqnarray}}
\newcommand{\nn}{\nonumber}

\numberwithin{equation}{section}

\begin{document}
{\begin{flushright}{APCTP Pre2022 - 012}\end{flushright}}

\title{$SU(4)_C \times SU(2)_L \times U(1)_R$ models with modular $A_4$ symmetry}

\author{Takaaki Nomura}
\email{nomura@scu.edu.cn}
\affiliation{College of Physics, Sichuan University, Chengdu 610065, China}

\author{Hiroshi Okada}
\email{hiroshi.okada@apctp.org}
\affiliation{Asia Pacific Center for Theoretical Physics (APCTP) - Headquarters San 31, Hyoja-dong,
Nam-gu, Pohang 790-784, Korea}
\affiliation{Department of Physics, Pohang University of Science and Technology, Pohang 37673, Republic of Korea}

\author{Yutaro Shoji}
\email{yutaro.shoji@mail.huji.ac.il}
\affiliation{Racah Institute of Physics, Hebrew University of Jerusalem, Jerusalem 91904, Israel}

\date{\today}

\begin{abstract}
{
We study $SU(4)_C \times SU(2)_L \times U(1)_R$ models with modular $A_4$ symmetry that provide unified description of the quark and lepton sector including the flavor structures.
The models are distinguished by the assignments of the modular weight on matter superfields.
We carry out numerical $\chi^2$ analysis and search for parameter sets that accomodate the experimental results.
We provide a benchmark point for each model to illustrate implications of our models.
}
\end{abstract}
\maketitle
\newpage

\section{Introduction}

Unified description of the quark and lepton sector is one of the biggest challenges in particle physics.
Grand unified theories (GUTs) are promising candidates of the quark-lepton unification that are based on a large gauge group such as SU(5), SO(10), $E_6$, etc, where the 
quarks and leptons are contained in multiplets of the unified gauge group.
Among many possibilities, an $SU(4)_C \times SU(2)_L \times U(1)_R$ gauge symmetry is an interesting choice to unify the 
quark and lepton sector since it is one of the minimal extensions of the standard model (SM)~\cite{Smirnov:1995jq}.
Interestingly, the scale of unification can be lower than the typical GUT scale, which gives a rich phenomenology~\cite{FileviezPerez:2022rbk,FileviezPerez:2021lkq,Faber:2018qon,FileviezPerez:2013zmv}.
For example, it is discussed that experimental anomalies in semi-leptonic $B$ meson decay can be addressed in the framework~\cite{FileviezPerez:2022rbk, FileviezPerez:2021lkq, Faber:2018qon}.

A challenge in this kind of models is the reproduction of the flavor structures.
One interesting approach is a modular flavor symmetry proposed in Ref.~\cite{Feruglio:2017spp,deAdelhartToorop:2011re}.
The structures of dimensionless couplings, especially the flavor structures of the Yukawa couplings,
are uniquely determined once we fix the modular weight, which is an additional quantum number associated with the modular symmetry. 
In this framework, we obtain more predictive models compared with traditional flavor models. A plethora of such scenarios can be found in literature; especially with a modular $A_4$ symmetry, see~\cite{Feruglio:2017spp, Criado:2018thu, Kobayashi:2018scp, Okada:2018yrn, Nomura:2019jxj, Okada:2019uoy, deAnda:2018ecu, Novichkov:2018yse, Nomura:2019yft, Okada:2019mjf,Ding:2019zxk, Nomura:2019lnr,Kobayashi:2019xvz,Asaka:2019vev,Zhang:2019ngf, Gui-JunDing:2019wap,Kobayashi:2019gtp,Nomura:2019xsb, Wang:2019xbo,Okada:2020dmb,Okada:2020rjb, Behera:2020lpd, Behera:2020sfe, Nomura:2020opk, Nomura:2020cog, Asaka:2020tmo, Okada:2020ukr, Nagao:2020snm, Okada:2020brs, Yao:2020qyy, Chen:2021zty, Kashav:2021zir, Okada:2021qdf, deMedeirosVarzielas:2021pug, Nomura:2021yjb, Hutauruk:2020xtk, Ding:2021eva, Nagao:2021rio, king, Okada:2021aoi, Nomura:2021pld, Kobayashi:2021pav, Dasgupta:2021ggp, Liu:2021gwa, Nomura:2022hxs, Otsuka:2022rak,Kobayashi:2021ajl, Chauhan:2022gkz, Kikuchi:2022pkd, Kobayashi:2022jvy, Gehrlein:2022nss, Almumin:2022rml,Kashav:2022kpk,Feruglio:2021dte}.
\footnote{For interested readers, we provide useful review references~\cite{Altarelli:2010gt, Ishimori:2010au, Ishimori:2012zz, Hernandez:2012ra, King:2013eh, King:2014nza, King:2017guk, Petcov:2017ggy, Kobayashi:2022moq}.}  
In these scenarios, the unification of the quark-lepton flavor is realized by applying both a unified gauge group and a modular flavor symmetry.
Moreover it is interesting to note that both the modular flavor symmetry and the unified gauge group can be originated from 
higher dimensional scenarios such as stringy ones.

In this paper, we consider supersymmetric models based on an $SU(4)_C \times SU(2)_L \times U(1)_R$ gauge symmetry and an $A_4$ modular flavor symmetry.
The models are distinguished by the assignments of the modular weight on matter superfields, which affect the structure of Yukawa couplings.
In these models,
the charged-lepton mass matrix and the down quark mass matrix share the same input parameters, as do the up quark mass matrix and the Dirac neutrino mass matrix.
Here, the active neutrino mass matrix is realized via the inverse seesaw mechanism~\cite{Mohapatra:1986bd, Wyler:1982dd} introducing left-handed neutral fermions.
The two mass matrices sharing the same input parameters differ by the way of linear combination as can be seen in Eq.~(\ref{eq:fermion-mass}).
It implies that fine-tuning of input parameters is inevitable in order to reproduce the observed quark and lepton masses and their mixings even in case where any flavor symmetry is not imposed.
Needless to say, the fine-tuning is much more stringent when we impose the flavor symmetry.
Since the input parameters are non-trivially related to the mixing parameters, an analytical analysis would be impossible in our case.
Thus, we carry out a numerical $\chi^2$ analysis searching for parameter sets that accommodate the measured values in quark, charged lepton and neutrino sectors.
For each model, we show a benchmark point (BP) with the smallest $\chi^2$ value among we found, and discuss which model is realistic.

Our manuscript is organized as follows.
{In Sec.~\ref{sect2}, we introduce our models based on  the $SU(4)_C \times SU(2)_L \times U(1)_R$ gauge symmetry and the $A_4$ modular flavor symmetry, and formulate mass matrices in the quark and lepton sectors.
In Sec.~\ref{sect3}, we perform numerical analysis and show benchmark points with minimum $\chi^2$ values for the observables associated with quarks, charged leptons and neutrinos.
 Finally, we conclude and discuss in Sec.~\ref{sec:conclusion}.}

\section{Models}
\label{sect2}

\begin{center} 
\begin{table}[tb]
\begin{tabular}{||c||c|c|c|c||c|c|c|c|c|c||}\hline
 &\multicolumn{10}{c||}{Matter superfields}  \\ \hline \hline
Fields  & ~$\hat{F}_{QL}$~ & ~$\hat{F}^c_u$~ & ~$\hat{F}^c_d$~ & ~$\hat N$~ & ~$\hat H_u$~ & ~$\hat H_d$~  & ~$\hat\Phi_u$~  & ~$\hat\Phi_d$~  & ~$\hat\chi_1$~ & ~$\hat\chi_2$~  \\\hline 
 $SU(4)_C$ & $\bm{4}$     & $\overline{\bm{4}}$ & $\overline{\bm{4}}$  & $\bm{1}$ & $\bm{1}$ & $\bm{1}$& $\bm{15}$ & $\bm{15}$ & $\bm{4}$ & $\bm{4}$    \\\hline 
  $SU(2)_L$ & $\bm{2}$     & $\bm{1}$  & $\bm{1}$  & $\bm{1}$ & $\bm{2}$ & $\bm{2}$& $\bm{2}$ & $\bm{2}$ & $\bm{1}$ & $\bm{1}$    \\\hline 
$U(1)_R$ & $0$  & $-\frac12$   & $\frac12$ & $0$  & $\frac12$  & -$\frac12$   & $\frac12$  & $-\frac12$     & $\frac12$ & $-\frac12$   \\\hline
 $A_4$ & $3$ & $r_{\rm singlets}$  & $r_{\rm singlets}$ & $3$ & $1$ & $1$ & $1$ & $1$ & $1$ & $1$ \\\hline
 $-k$ & $-2$ & $k_u$ & $k_d$ & $-2$ & $0$ & $0$& $0$ & $0$ & $0$  & $0$   \\\hline
\end{tabular}
\caption{Matter chiral superfields and their charge assignments under $SU(4)_C\times SU(2)_L\times U(1)_R\times A_4$, where $-k$ is the number of modular weight.}
\label{tab:1}
\end{table}
\end{center}

\begin{center} 
\begin{table}[tb]
\begin{tabular}{|c|c|c|c|c|}\hline
~&~ model (I) ~&~ model (II) ~&~ model (III) ~&~ model (IV)~ \\ \hline
$r_{\rm singlets}$ & $\{1, 1'', 1' \}$ & $\{1, 1'', 1' \}$ & $\{1, 1'', 1' \}$ & $\{1, 1', 1' \}$ \\
$k_u$ & $-2$ & $0$ & $-4$ & $-4$ \\
$k_d$ & $0$ & $-4$ & $0$ & $-4$ \\ \hline  
\end{tabular}
\caption{Assignments of $r_{\rm singlets}$ and $k_{u(d)}$ for each model. The assignment of $r_{\rm singlets}$ is flavor-dependent.}
\label{tab:2}
\end{table}
\end{center}

In this section, we provide the setup of our supersymmetric models based on an $SU(4)_C \times SU(2)_L \times U(1)_R$ gauge symmetry and an $A_4$ modular flavor symmetry.
For the quark and lepton sectors, we introduce chiral superfields, $\hat{F}_{QL}$, $\hat{F}^c_u$, $\hat{F}^c_d$ and $\hat N$, that respectively belong to $(\bm{4}, \bm{2}, 0)$, $(\overline{\bm{4}}, \bm{1}, -1/2)$, $(\overline{\bm{4}}, \bm{1} , 1/2)$ and $(\bm{1}, \bm{1}, 0)$ representations under $SU(4)_C \times SU(2)_L \times U(1)_R$.
These $SU(4)$ multiplets can be written in terms of the SM gauge representations;
\begin{equation}
F_{QL} = \left( \begin{array}{c} Q_L \\ L_L \end{array} \right), \quad
F^c_{u} = \left( \begin{array}{cc} u^c_R & \nu^c_R \end{array} \right), \quad
F^c_{d} = \left( \begin{array}{cc} d^c_R & e^c_R \end{array} \right).
\end{equation}
For the Higgs sector, we introduce chiral superfields, $\hat{H}_{u(d)}$, $\hat{\Phi}_{u(d)}$ and $\hat{\chi}_{1(2)}$, that respectively belong to $(\bm{1}, \bm{2}, 1/2(-1/2))$, $(\bm{15}, \bm{2}, 1/2(-1/2))$ and $(\bm{4}, \bm{1}, 1/2(-1/2))$ representations.
Note that we introduced two sets of chiral super fields with opposite $U(1)_R$ charge in the Higgs sector to cancel gauge anomalies.
These $SU(4)$ multiplets can be written in terms of the SM gauge representations;
\begin{align}
\Phi_{u(d)} = \left( \begin{array}{cc} \Phi_{8u(8d)} & \Phi_{3u(3d)} \\ \Phi_{4u(4d)} & 0 \end{array} \right) + T_4 H'_{u(d)}, \quad
\chi_1 = \left( \begin{array}{c} \chi_u \\ \chi^0_u \end{array} \right), \quad
\chi_2 = \left( \begin{array}{c} \chi_d \\ \chi^-_d \end{array} \right),
\end{align}
where $T_4 = 1/(2\sqrt6){\rm diag}(1,1,1,-3)$ is a diagonal $SU(4)_C$ generator, and $\Phi_{8u(8d)}$, $\Phi_{3u(3d)}$, $\Phi_{4u(4d)}$, $H'_{u(d)}$, $\chi_{u(d)}$, $\chi^0_u$ and $\chi^-_{d}$ belong to representations of $(8,2,1/2(-1/2))$
, $(\bar{3},2,-1/6(-7/6))$, $(3,2,7/6(1/6))$, $(1,2,1/2(-1/2))$, $(3,1,2/3(-1/3))$, $(1,1,0)$ and $(1,1,-1)$ under $SU(3)_C \times SU(2)_L \times U(1)_Y$.
In our scenario, we assume electrically neutral components of the scalar fields in the Higgs sector develop vacuum expectation values (VEVs); 
$\langle H_{u(d)} \rangle = v_{u(d)}/\sqrt{2}$, $\langle H'_{u(d)} \rangle = v'_{u(d)}/\sqrt{2}$  and $\langle \chi^0_u \rangle = v_\chi/\sqrt2$.
We parameterize the VEVs as
\begin{align}
& v_1 = v \sin \beta, \ v_2 = v \cos \beta, \nonumber \\
&  v_{u} = v_1 \cos a, \ v_d = v_2 \cos b, \ v'_u = v_1 \sin a, \ v'_d  = v_2 \sin b, 
\end{align}
where $v \simeq 246$ GeV and we chose $0 \leq \{\beta, a, b\} \leq \pi/2$ without loss of generality.
The $SU(4)_C$ gauge symmetry is broken by the VEV of $\chi^0_u$ and the hypercharge, $Y$, is given by
\begin{equation}
Y = R + \frac{\sqrt6}{3} T_4,
\end{equation}
where $R$ is the $U(1)_R$ charge.
In Table~\ref{tab:1} and \ref{tab:2}, we summarize the field contents and the charge assignments,
where different models are distinguished by the modular weight and the $A_4$ singlet representation of $\hat{F}^c_u$ and $\hat{F}^c_d$.
Note that model (IV) satisfies the criterion of scenarios in which the modular flavor symmetry is obtained from a higher-dimensional 
theory and the modular forms are understood as wave functions in extra dimensions~\cite{Kikuchi:2022txy}.

In the following, we construct a superpotential relevant to produce fermion masses, where each term of the superpotential should 
have zero modular weight to be invariant under the $A_4$ modular symmetry.

\subsection{Mass matrices for quarks and charged leptons}

In our setup, the superpotential relevant for the generation of the SM fermion masses is given by
\begin{align}
w_{QL} = Y_1 \hat{F}^c_u\hat F_{QL} \hat H_u + Y_2 \hat{F}^c_u \hat \Phi_{u} \hat F_{QL} + Y_3 \hat{F}^c_d \hat F_{QL} \hat H_d + Y_4 \hat{F}^c_d \hat \Phi_{d} \hat F_{QL},
\end{align}
where the structures of Yukawa coupling $Y_{i=1-4}$ depends on models and we ommit the flavor indices.
Expanding the multiplets by their components, we obtain
\begin{align}
\label{eq:Yukawa1}
w_{QL} =  & Y_1 \left[ \hat{u}^c_R \hat{Q}_L \hat{H}_u + \hat{\nu}^c_R \hat{L}_L \hat{H}_u \right] 
+ Y_3 \left[ \hat{d}^c_R \hat{Q}_L \hat{H}_d + \hat{e}^c_R \hat{L}_L \hat{H}_d \right]  \nonumber \\
& + Y_2 \left[ \hat{u}^c_R \hat{\Phi}_{u8} \hat{Q}_L + \hat{u}^c_R \hat{\Phi}_{u3} \hat{L}_L + \hat{\nu}^c_R \hat{\Phi}_{u4} \hat{Q}_L 
+ \frac{1}{2\sqrt{6}} \hat{u}^c_R \hat{Q}_L \hat{H}'_u - \frac{3}{2\sqrt{6}} \hat{\nu}^c_R \hat{L}_L \hat{H}'_u \right] \nonumber \\
& + Y_4 \left[ \hat{d}^c_R \hat{\Phi}_{d8} \hat{Q}_L + \hat{d}^c_R \hat{\Phi}_{d3} \hat{L}_L + \hat{e}^c_R \hat{\Phi}_{d4} \hat{Q}_L 
+ \frac{1}{2\sqrt{6}} \hat{d}^c_R \hat{Q}_L \hat{H}'_d - \frac{3}{2\sqrt{6}} \hat{e}^c_R \hat{L}_L \hat{H}'_d \right].
\end{align} 
After the spontaneous symmetry breaking, we obtain the mass terms for the SM quarks and the charged leptons as
\begin{align}
& -\mathcal L_M = \overline{u_R} M_u u_L + \overline{d_R} M_d d_L + \overline{e_R} M_e e_L + h.c., \nonumber \\ 
& M_u = \frac{v_u}{\sqrt{2}} Y_1 + \frac{v'_u}{4 \sqrt{3}} Y_2, \quad
M_d = \frac{v_d}{\sqrt{2}} Y_3 + \frac{3v'_d}{4 \sqrt{3}} Y_4, \quad
M_e = \frac{v_d}{\sqrt{3}} Y_3 - \frac{3v'_d}{4 \sqrt{3}} Y_4, 
\label{eq:fermion-mass}
\end{align}
where $\{u,d,e\}$ denote the SM fermions in the flavor basis.
In the discussion below, we focus on the SM fermion masses and the neutrino masses, and do not discuss those of the
superpartners and the extra particle contents assuming they are heavy enough.

In the following, we summarize the structures of Yukawa couplings $Y_{i=1-4}$ for each model. \\
{\bf Model (I) } \\
In this model, the Yukawa couplings are given as follows:
\begin{align}
& Y_1=
\left[\begin{array}{ccc}
a_1 & 0 & 0 \\ 
0 & a_2 & 0  \\ 
0 & 0 & a_3  \\ 
\end{array}\right]
\left[\begin{array}{ccc}
y^{(4)}_1 &  y^{(4)}_3 & y^{(4)}_2 \\ 
 y^{(4)}_2 &  y^{(4)}_1 & y^{(4)}_3  \\ 
 y^{(4)}_3 &  y^{(4)}_2 & y^{(4)}_1  \\
\end{array}\right], \quad
Y_2=
\left[\begin{array}{ccc}
b_1 & 0 & 0 \\ 
0 & b_2 & 0  \\ 
0 & 0 & b_3  \\ 
\end{array}\right]
\left[\begin{array}{ccc}
y^{(4)}_1 &  y^{(4)}_3 & y^{(4)}_2 \\ 
 y^{(4)}_2 &  y^{(4)}_1 & y^{(4)}_3  \\ 
 y^{(4)}_3 &  y^{(4)}_2 & y^{(4)}_1  \\
\end{array}\right], \nonumber \\
& Y_3=
\left[\begin{array}{ccc}
c_1 & 0 & 0 \\ 
0 & c_2 & 0  \\ 
0 & 0 & c_3  \\ 
\end{array}\right]
\left[\begin{array}{ccc}
y_1 &  y_3 & y_2 \\ 
 y_2 &  y_1 & y_3  \\ 
 y_3 &  y_2 & y_1  \\
\end{array}\right], \quad
Y_4=
\left[\begin{array}{ccc}
d_1 & 0 & 0 \\ 
0 & d_2 & 0  \\ 
0 & 0 & d_3  \\ 
\end{array}\right]
\left[\begin{array}{ccc}
y_1 &  y_3 & y_2 \\ 
 y_2 &  y_1 & y_3  \\ 
 y_3 &  y_2 & y_1  \\
\end{array}\right], 
\end{align}
where $\{a_i, b_i, c_i, d_i\} (i=1-3)$  are complex parameters. \\
%
%
{\bf Model (II)}\\
In this model, $Y_1$ and $Y_2$ are the same as those of model (I) while $Y_3$ and $Y_4$ are written in terms of weight 6 modular forms as 
\begin{align}
& Y_3=
\left[\begin{array}{ccc}
c_1 & 0 & 0 \\ 
0 & c_2 & 0  \\ 
0 & 0 & c_3  \\ 
\end{array}\right]
\left[\begin{array}{ccc}
y^{(6)}_1 &  y^{(6)}_3 & y^{(6)}_2 \\ 
 y^{(6)}_2 &  y^{(6)}_1 & y^{(6)}_3  \\ 
 y^{(6)}_3 &  y^{(6)}_2 & y^{(6)}_1  \\
\end{array}\right]
+
\left[\begin{array}{ccc}
c_4 & 0 & 0 \\ 
0 & c_5 & 0  \\ 
0 & 0 & c_6  \\ 
\end{array}\right]
\left[\begin{array}{ccc}
y^{'(6)}_1 &  y^{'(6)}_3 & y^{'(6)}_2 \\ 
 y^{'(6)}_2 &  y^{'(6)}_1 & y^{'(6)}_3  \\ 
 y^{'(6)}_3 &  y^{'(6)}_2 & y^{'(6)}_1  \\
\end{array}\right], \nonumber \\
& Y_4=
\left[\begin{array}{ccc}
d_1 & 0 & 0 \\ 
0 & d_2 & 0  \\ 
0 & 0 & d_3  \\ 
\end{array}\right]
\left[\begin{array}{ccc}
y^{(6)}_1 &  y^{(6)}_3 & y^{(6)}_2 \\ 
 y^{(6)}_2 &  y^{(6)}_1 & y^{(6)}_3  \\ 
 y^{(6)}_3 &  y^{(6)}_2 & y^{(6)}_1  \\
\end{array}\right]
+
\left[\begin{array}{ccc}
d_4 & 0 & 0 \\ 
0 & d_5 & 0  \\ 
0 & 0 & d_6  \\ 
\end{array}\right]
\left[\begin{array}{ccc}
y^{'(6)}_1 &  y^{'(6)}_3 & y^{'(6)}_2 \\ 
 y^{'(6)}_2 &  y^{'(6)}_1 & y^{'(6)}_3  \\ 
 y^{'(6)}_3 &  y^{'(6)}_2 & y^{'(6)}_1  \\
 \end{array}\right].
\end{align}
{\bf Model (III)}\\
In this model, $Y_{3}$ and $Y_4$ are the same as those of model (I) while $Y_1$ and $Y_2$ are written in terms of weight 6 modular forms as
\begin{align}
& Y_1=
\left[\begin{array}{ccc}
a_1 & 0 & 0 \\ 
0 & a_2 & 0  \\ 
0 & 0 & a_3  \\ 
\end{array}\right]
\left[\begin{array}{ccc}
y^{(6)}_1 &  y^{(6)}_3 & y^{(6)}_2 \\ 
 y^{(6)}_2 &  y^{(6)}_1 & y^{(6)}_3  \\ 
 y^{(6)}_3 &  y^{(6)}_2 & y^{(6)}_1  \\
\end{array}\right]
+
\left[\begin{array}{ccc}
a_4 & 0 & 0 \\ 
0 & a_5 & 0  \\ 
0 & 0 & a_6  \\ 
\end{array}\right]
\left[\begin{array}{ccc}
y^{'(6)}_1 &  y^{'(6)}_3 & y^{'(6)}_2 \\ 
 y^{'(6)}_2 &  y^{'(6)}_1 & y^{'(6)}_3  \\ 
 y^{'(6)}_3 &  y^{'(6)}_2 & y^{'(6)}_1  \\
\end{array}\right], \nonumber \\
& Y_2=
\left[\begin{array}{ccc}
b_1 & 0 & 0 \\ 
0 & b_2 & 0  \\ 
0 & 0 & b_3  \\ 
\end{array}\right]
\left[\begin{array}{ccc}
y^{(6)}_1 &  y^{(6)}_3 & y^{(6)}_2 \\ 
 y^{(6)}_2 &  y^{(6)}_1 & y^{(6)}_3  \\ 
 y^{(6)}_3 &  y^{(6)}_2 & y^{(6)}_1  \\
\end{array}\right]
+
\left[\begin{array}{ccc}
b_4 & 0 & 0 \\ 
0 & b_5 & 0  \\ 
0 & 0 & b_6  \\ 
\end{array}\right]
\left[\begin{array}{ccc}
y^{'(6)}_1 &  y^{'(6)}_3 & y^{'(6)}_2 \\ 
 y^{'(6)}_2 &  y^{'(6)}_1 & y^{'(6)}_3  \\ 
 y^{'(6)}_3 &  y^{'(6)}_2 & y^{'(6)}_1  \\
 \end{array}\right].
\end{align}
{\bf Model (IV)}\\
In this model, all the Yukawa couplings, $Y_{i=1-4}$, are written in terms of weight 6 modular forms such that
\begin{align}
& Y_1=
\left[\begin{array}{ccc}
a_1 & 0 & 0 \\ 
0 & a_2 & 0  \\ 
0 & 0 & a_3  \\ 
\end{array}\right]
\left[\begin{array}{ccc}
y^{(6)}_1 &  y^{(6)}_3 & y^{(6)}_2 \\ 
 y^{(6)}_3 &  y^{(6)}_2 & y^{(6)}_1  \\ 
 y^{(6)}_3 &  y^{(6)}_2 & y^{(6)}_1  \\
\end{array}\right]
+
\left[\begin{array}{ccc}
a_4 & 0 & 0 \\ 
0 & a_5 & 0  \\ 
0 & 0 & a_6  \\ 
\end{array}\right]
\left[\begin{array}{ccc}
y^{'(6)}_1 &  y^{'(6)}_3 & y^{'(6)}_2 \\ 
 y^{'(6)}_3 &  y^{'(6)}_2 & y^{'(6)}_1  \\ 
 y^{'(6)}_3 &  y^{'(6)}_2 & y^{'(6)}_1  \\
\end{array}\right], \nonumber \\
& Y_2=
\left[\begin{array}{ccc}
b_1 & 0 & 0 \\ 
0 & b_2 & 0  \\ 
0 & 0 & b_3  \\ 
\end{array}\right]
\left[\begin{array}{ccc}
y^{(6)}_1 &  y^{(6)}_3 & y^{(6)}_2 \\ 
 y^{(6)}_3 &  y^{(6)}_2 & y^{(6)}_1  \\ 
 y^{(6)}_3 &  y^{(6)}_2 & y^{(6)}_1  \\
\end{array}\right]
+
\left[\begin{array}{ccc}
b_4 & 0 & 0 \\ 
0 & b_5 & 0  \\ 
0 & 0 & b_6  \\ 
\end{array}\right]
\left[\begin{array}{ccc}
y^{'(6)}_1 &  y^{'(6)}_3 & y^{'(6)}_2 \\ 
 y^{'(6)}_3 &  y^{'(6)}_2 & y^{'(6)}_1  \\ 
 y^{'(6)}_3 &  y^{'(6)}_2 & y^{'(6)}_1  \\
\end{array}\right], \nonumber \\
& Y_3=
\left[\begin{array}{ccc}
c_1 & 0 & 0 \\ 
0 & c_2 & 0  \\ 
0 & 0 & c_3  \\ 
\end{array}\right]
\left[\begin{array}{ccc}
y^{(6)}_1 &  y^{(6)}_3 & y^{(6)}_2 \\ 
 y^{(6)}_3 &  y^{(6)}_2 & y^{(6)}_1  \\ 
 y^{(6)}_3 &  y^{(6)}_2 & y^{(6)}_1  \\
\end{array}\right]
+
\left[\begin{array}{ccc}
c_4 & 0 & 0 \\ 
0 & c_5 & 0  \\ 
0 & 0 & c_6  \\ 
\end{array}\right]
\left[\begin{array}{ccc}
y^{'(6)}_1 &  y^{'(6)}_3 & y^{'(6)}_2 \\ 
 y^{'(6)}_3 &  y^{'(6)}_2 & y^{'(6)}_1  \\ 
 y^{'(6)}_3 &  y^{'(6)}_2 & y^{'(6)}_1  \\
\end{array}\right], \nonumber \\
& Y_4=
\left[\begin{array}{ccc}
d_1 & 0 & 0 \\ 
0 & d_2 & 0  \\ 
0 & 0 & d_3  \\ 
\end{array}\right]
\left[\begin{array}{ccc}
y^{(6)}_1 &  y^{(6)}_3 & y^{(6)}_2 \\ 
 y^{(6)}_3 &  y^{(6)}_2 & y^{(6)}_1  \\ 
 y^{(6)}_3 &  y^{(6)}_2 & y^{(6)}_1  \\
\end{array}\right]
+
\left[\begin{array}{ccc}
d_4 & 0 & 0 \\ 
0 & d_5 & 0  \\ 
0 & 0 & d_6  \\ 
\end{array}\right]
\left[\begin{array}{ccc}
y^{'(6)}_1 &  y^{'(6)}_3 & y^{'(6)}_2 \\ 
 y^{'(6)}_3 &  y^{'(6)}_2 & y^{'(6)}_1  \\ 
 y^{'(6)}_3 &  y^{'(6)}_2 & y^{'(6)}_1  \\
\end{array}\right].
\end{align}

Substituting these Yukawa couplings into Eq.~\eqref{eq:fermion-mass} and diagonalizing the mass matrices, we obtain the fermion masses and the mixings.
Quark mass matrices $M_u$ and $M_d$ are diagonalized by two unitary matrices as $D_d = V^\dagger_{d_L} M_d V_{d_R}$ and $D_u = V^\dagger_{u_L} M_{u} V_{u_R}$,
where $D_d \equiv {\rm}(m_d, m_s, m_b)$ and $D_u \equiv {\rm}(m_u, m_c, m_t)$ represent mass eigenvalues corresponding to the SM quark masses.
We then get the CKM matrix as
\begin{equation}
V_{CKM} = V^\dagger_{u_L} V_{d_L}.
\end{equation}
Similarly charged lepton mass matrix $M_\ell$ is diagonalized by two unitary matrices as $D_\ell = V^\dagger_{\ell_L} M_\ell V_{\ell_R}$,
where $D_\ell \equiv {\rm dig}(m_e, m_\mu, m_\tau)$ is mass eigenvalues corresponding to the SM charged lepton masses.

\subsection{Neutrino mass matrix}
\label{subsec:neutrino}
In our models, the active neutrino masses arise from the inverse seesaw mechanism.
The Dirac type mass terms, $m_D^\dag \overline{\nu_R} \nu_L + h.c.$, are obtained from the terms with coupling $Y_1$ and $Y_2$ in Eq.~\eqref{eq:Yukawa1}. 
Additional terms in the superpotential that are relevant to the inverse seesaw mechanism are given by
\begin{equation}
w_v = Y_5 F^c_u \chi_1 N + \frac12 \mu \bar N^c N,
\end{equation}
where $Y_5$ and $\mu$ are given by modular forms.
The structure of $Y_5$ is written by 
\begin{align}
& Y_5=
\left[\begin{array}{ccc}
e_1 & 0 & 0 \\ 
0 & e_2 & 0  \\ 
0 & 0 & e_3  \\ 
\end{array}\right]
\left[\begin{array}{ccc}
y^{(4)}_1 &  y^{(4)}_3 & y^{(4)}_2 \\ 
 y^{(4)}_2 &  y^{(4)}_1 & y^{(4)}_3  \\ 
 y^{(4)}_3 &  y^{(4)}_2 & y^{(4)}_1  \\
\end{array}\right] \quad \text{[model (I), (II)]}, \\
& Y_5=
\left[\begin{array}{ccc}
e_1 & 0 & 0 \\ 
0 & e_2 & 0  \\ 
0 & 0 & e_3  \\ 
\end{array}\right]
\left[\begin{array}{ccc}
y^{(6)}_1 &  y^{(6)}_3 & y^{(6)}_2 \\ 
 y^{(6)}_2 &  y^{(6)}_1 & y^{(6)}_3  \\ 
 y^{(6)}_3 &  y^{(6)}_2 & y^{(6)}_1  \\
\end{array}\right]
+
\left[\begin{array}{ccc}
e_4 & 0 & 0 \\ 
0 & e_5 & 0  \\ 
0 & 0 & e_6  \\ 
\end{array}\right]
\left[\begin{array}{ccc}
y^{'(6)}_1 &  y^{'(6)}_3 & y^{'(6)}_2 \\ 
 y^{'(6)}_2 &  y^{'(6)}_1 & y^{'(6)}_3 \\ 
 y^{'(6)}_3 &  y^{'(6)}_2 & y^{'(6)}_1  \\
\end{array}\right] \quad \text{[model (III)]}, \nonumber\\
& Y_5=
\left[\begin{array}{ccc}
e_1 & 0 & 0 \\ 
0 & e_2 & 0  \\ 
0 & 0 & e_3  \\ 
\end{array}\right]
\left[\begin{array}{ccc}
y^{(6)}_1 &  y^{(6)}_3 & y^{(6)}_2 \\ 
 y^{(6)}_3 &  y^{(6)}_2 & y^{(6)}_1  \\ 
 y^{(6)}_3 &  y^{(6)}_2 & y^{(6)}_1  \\
\end{array}\right]
+
\left[\begin{array}{ccc}
e_4 & 0 & 0 \\ 
0 & e_5 & 0  \\ 
0 & 0 & e_6  \\ 
\end{array}\right]
\left[\begin{array}{ccc}
y^{'(6)}_1 &  y^{'(6)}_3 & y^{'(6)}_2 \\ 
 y^{'(6)}_3 &  y^{'(6)}_2 & y^{'(6)}_1  \\ 
 y^{'(6)}_3 &  y^{'(6)}_2 & y^{'(6)}_1  \\
\end{array}\right] \quad \text{[model (IV)]}. \nonumber
\end{align}
The structure of $\mu$ is common in all of our models and is given by
\begin{equation}
\mu = \mu_0 \left\{\frac13 \left[ \begin{array}{ccc}
2 y^{(4)}_1 & - y^{(4)}_3 & -y^{(4)}_2 \\ 
- y^{(4)}_3 &  2y^{(4)}_2 & -y^{(4)}_1  \\ 
- y^{(4)}_2 &  -y^{(4)}_1 & 2y^{(4)}_3  \\
\end{array}\right] 
+ \alpha_N \left[ \begin{array}{ccc} 1 & 0 & 0 \\ 0 & 0 & 1 \\ 0 & 1& 0 \end{array} \right]
+ \beta_N \left[ \begin{array}{ccc} 0 & 0 & 1 \\ 0 & 1 & 0 \\ 1 & 0& 0 \end{array} \right]
 \right\}.
\end{equation}
After the spontaneous symmetry breaking, we obtain a $9\times9$ neutral fermion mass matrix, in the basis of $(\nu^c_L, \nu_R, N_L)^T$, such that
\begin{equation}
M_N = \left[ \begin{array}{ccc} 0 & m_D & 0 \\ m_D^T & 0 & M^T \\ 0 & M & \mu \end{array} \right],
\end{equation}
where each element corresponds to a $3\times3$ matrix.
The $3\times 3$ matrices, $m_D$ and $M$, are written by
\begin{align}
m_D^T = \frac{v_u}{\sqrt2} Y_1 - \frac{3 v'_u}{4\sqrt{3}} Y_2, \quad  M^T = \frac{v_\chi}{\sqrt2} Y_5. 
\end{align}
Then the active neutrino mass matrix can be approximately obtained as 
\begin{equation}
m_\nu \simeq m_D M^{-1} \mu (M^T)^{-1} m_D^T,
\end{equation}
where a hierarchy of mass scale $\mu \ll m_D \lesssim M$ is expected.
Then we diagonalize the neutrino mass matrix  by unitary matrix $U_\nu$; $D_\nu = U^T_\nu m_\nu U_\nu$, where $D_\nu \equiv {\rm diag}(m_1,m_2,m_3)$ represents mass eigenvalues.
The PMNS matrix is defined by $U_{\rm PMNS} \equiv V^\dagger_L U_\nu$, where $V_L^\dagger$ comes from diagonalization of the charged lepton mass matrix. 
We parametrize $U_{\rm PMNS}$ in terms of three mixing angles $\theta_{ij} (i,j=1,2,3; i < j)$, one CP violating Dirac phase $\delta_{CP}$,
and two Majorana phases $\{\alpha_{21}, \alpha_{32}\}$ as follows:
\begin{equation}
U_{\rm PMNS}= 
\begin{pmatrix} c_{12} c_{13} & s_{12} c_{13} & s_{13} e^{-i \delta_{CP}} \\ 
-s_{12} c_{23} - c_{12} s_{23} s_{13} e^{i \delta_{CP}} & c_{12} c_{23} - s_{12} s_{23} s_{13} e^{i \delta_{CP}} & s_{23} c_{13} \\
s_{12} s_{23} - c_{12} c_{23} s_{13} e^{i \delta_{CP}} & -c_{12} s_{23} - s_{12} c_{23} s_{13} e^{i \delta_{CP}} & c_{23} c_{13} 
\end{pmatrix}
\begin{pmatrix} 1 & 0 & 0 \\ 0 & e^{i \frac{\alpha_{21}}{2}} & 0 \\ 0 & 0 & e^{i \frac{\alpha_{31}}{2}} \end{pmatrix},
\end{equation}
where $c_{ij}$ and $s_{ij}$ respectively stand for $\cos \theta_{ij}$ and $\sin \theta_{ij}$. 
Then, {each mixing angle} can be written in terms of components of $U_{\mathrm{PMNS}}$:
\begin{align}
\sin^2\theta_{13}=|(U_{\mathrm{PMNS}})_{13}|^2,\quad 
\sin^2\theta_{23}=\frac{|(U_{\mathrm{PMNS}})_{23}|^2}{1-|(U_{\mathrm{PMNS}})_{13}|^2},\quad 
\sin^2\theta_{12}=\frac{|(U_{\mathrm{PMNS}})_{12}|^2}{1-|(U_{\mathrm{PMNS}})_{13}|^2}.
\end{align}
In addition, we compute the Jarlskog invariant and $\delta_{CP}$ derived from PMNS matrix elements $U_{\alpha i}$:
\begin{equation}
J_{CP} = \text{Im} [U_{e1} U_{\mu 2} U_{e 2}^* U_{\mu 1}^*] = s_{23} c_{23} s_{12} c_{12} s_{13} c^2_{13} \sin \delta_{CP}.
\end{equation}
We also estimate the Majorana phases in terms of other invariants $I_1$ and $I_2$:
\begin{equation}
I_1 = \text{Im}[U^*_{e1} U_{e2}] = c_{12} s_{12} c_{13}^2 \sin \left( \frac{\alpha_{21}}{2} \right), \
I_2 = \text{Im}[U^*_{e1} U_{e3}] = c_{12} s_{13} c_{13} \sin \left( \frac{\alpha_{31}}{2} - \delta_{CP} \right).
\end{equation}
For neutrino masses, we write square differences as
\begin{align}
(\mathrm{NO}):\  \Delta m_{\rm atm}^2 =m_3^2 - m_1^2,
\quad
(\mathrm{IO}):\    \Delta m_{\rm atm}^2 =m_2^2 - m_3^2,
 \end{align}
where $\Delta m_{\rm atm}^2$ is the atmospheric neutrino mass square difference, and NO and IO represent the normal ordering and the inverted ordering, respectively. 
Then the solar mass square difference is given by:
\begin{align}
\Delta m_{\rm sol}^2=m_2^2 - m_1^2,
 \end{align}
 which will be compared with the observed value in our $\chi^2$ analysis.
Moreover, the effective mass for the neutrinoless double beta decay is obtained as
\begin{align}
\langle m_{ee}\rangle=  |m_1 \cos^2\theta_{12} \cos^2\theta_{13}+m_2 \sin^2\theta_{12} \cos^2\theta_{13}e^{i\alpha_{21}}+m_3 \sin^2\theta_{13}e^{i(\alpha_{31}-2\delta_{CP})}|,
\end{align}
where its value may be tested at the KamLAND-Zen experiment in future~\cite{KamLAND-Zen:2016pfg}. 
We adopt the neutrino experimental data in NuFit5.0~\cite{Esteban:2018azc} in carrying out numerical $\chi^2$ analysis. 

\section{Numerical analysis}
\label{sect3}

In this section, we perform numerical analysis searching for parameters that can accommodate all the experimental data of the quark and lepton masses and mixings including neutrinos.
The relevant free parameters are 
\begin{align}
& \{\tau, v_\chi, \sin \beta, \sin a, \sin b, a_i, b_j, c_k, d_l, e_m, \alpha_N, \beta_N \}, 
\end{align}
where indecies $\{i,j,k,l,m\}$ run from 1 to their maximal values given by
\begin{align}
&  \{i_{\rm max}, j_{\rm max}, k_{\rm max}, l_{\rm max}, m_{\rm max} \} =\{[3,3,6,6],[3,3,6,6],[3,6,3,6],[3,6,3,6],[3,3,6,6] \},  \nonumber \\
& \text{for model [(I), (II), (III), (IV)]}. \nonumber
\end{align}
We then minimize $\chi^2$ using these free parameters. 
The relevant observables are summarized as~\footnote{The $|V_{ub}|$, $|V_{cb}|$ and $|V_{ub}/V_{cb}|$ are observed independently and the value of $|V_{ub}/V_{cb}|$ can be deviated from $1\sigma$ region even if $|V_{ub}|$ and $|V_{cb}|$ values are within $1\sigma$. We thus compare our results with these three values although they are not independent theoretically. 
Note also that we do not consider the other ratios of the CKM elements since they are within $1\sigma$ level if the corresponding CKM elements are within 1$\sigma$.
In addition we are not considering the other component of CKM matrix such as $|V_{ts}|$ etc. because if the elements we included in our analysis satisfy observed range the rest ones are also within observed range 
due to the unitarity of the CKM matrix~\cite{ParticleDataGroup:2020ssz}.} 
\begin{align}
& \bigl\{m_e, m_\mu, m_\tau, m_u, m_c, m_t, m_d, m_s, m_b, |V_{us}|, |V_{cb}|, |V_{ub}|, \nonumber \\
& \ |V_{td}|,|V_{ub}/V_{cb}|, \delta_{CP}^{CKM}, \sin \theta_{23}, \sin \theta_{13}, \sin \theta_{12}, \Delta m^2_{\rm sol} \bigr\},
\end{align}
where $\delta_{CP}^{CKM}$ is the CP-violating phase in the CKM matrix, 
we use $\Delta m^2_{\rm atm}$ as an input value to fix $\mu_0$, and we do not include the neutrino CP phase, $\delta_{CP}$, in $\chi^2$.
Thus our $\chi^2$ corresponds to the case of 19 degrees of freedom.
Note also that, in this paper, we focus on normal mass ordering case for simplicity.

We optimized $\chi^2$ using the Metropolis Monte Carlo method together with the steepest decent method starting from random initial parameters.
We found model (I) cannot fit the measured values due to the smallness of the number of free parameters, where we obtained a minimum value of $\chi^2\simeq385$.
For the other models, we found parameter sets that can accommodate all the measured values.
In Tables~\ref{tab:3}, \ref{tab:4} and \ref{tab:5}, we provide our benchmark points (BPs) for model (II), (III) and (IV), respectively. 
We also show the input parameters and some outputs such as CKM elements and mixings associated with neutrino.  
In model (II) and (III), we found BPs with modulus $\tau$ close to fixed point $\tau = i$
while we have obtained $\tau$ closed to the fixed point $\tau = i \times \infty $ for the BP of model (IV).
However, $\chi^2$ is highly dependent on ${\rm Im}\tau$ and we could not find any BPs exactly on the fixed point. For model (III), we found other BPs having similar $\chi^2$ and all of them have ${\rm Im}\tau\sim 1.05$ even though the initial conditions for minimization are different. It implies that ${\rm Im}\tau$ effectively determines a combination of observables almost independently of the other parameters.

\begin{center} 
\begin{table}[tb]
\begin{tabular}{|c|c|}\hline 
$\tau$ & $0.0103+1.03 \, i$ \\ \hline
$v_\chi/ [{\rm GeV}]$ & $4.96 \times 10^{12}$ \\ \hline
$[\sin \beta, \sin a, \sin b]$ & $[0.9997, 0.9728, 0.8817]$ \\ \hline
$[a_1, a_2, a_3]$ & $[-0.0763-0.0106 \, i, 2.18+0.154 \, i, 0.509+0.00479 \, i]$ \\ \hline
$[b_1, b_2, b_3]$ & $[0.0892+0.0124 \, i, 1.03+0.0883 \, i, -0.587+0.00125 \, i]$ \\ \hline
$[c_1, c_2, c_3]$ & $[(-1.73+7.43 \, i) \times 10^{-4}, -0.144+0.223 \, i, -0.0125+0.0105 \, i]$ \\ \hline
$[c_4, c_5, c_6]$ & $[-(2.62+18.1 \, i)\times 10^{-4}, -0.0107+0.0152 \, i, 0.00173+0.160 \, i]$ \\ \hline
$[d_1, d_2, d_3]$ & $[(4.55-11.2 \, i)\times 10^{-4}, 0.0167+0.185 \, i, -0.00959-0.0380 \, i ]$ \\ \hline
$[d_4, d_5, d_6]$ & $[-(4.12+4.24 \, i)\times 10^{-4}, 0.0431+0.0112 \, i, 0.00100+0.582 \, i]$ \\ \hline
$[e_1, e_2, e_3]$ & $[-0.129+0.714 \, i, -0.000168-7.05 \, i, 0.571+0.00752 \, i]$ \\ \hline
$[\alpha_N, \beta_N]$ & $[0.0169+0.831 \, i, 0.00744-0.00657 \, i]$ \\ \hline \hline
$\bigl[ |V_{us}|,|V_{cb}|,|V_{ub}| \bigr]$ & $[0.224, 0.0420,0.00327]$ \\ \hline
$\bigl[ |V_{td}|,|V_{ub}/V_{cb}| \bigr]$ & $[0.00856, 0.0778]$ \\ \hline
$\delta_{CP}^{CKM}$/[deg] & 64.8 \\ \hline 
$[\sin^2 \theta_{12}, \sin^2 \theta_{23}, \sin^2 \theta_{13}]$ & $[0.304, 0.570, 0.0222]$ \\ \hline
$[\delta_{CP}, \alpha_1, \alpha_2]$/[deg] & $[273, 148, 32.6]$ \\ \hline 
$\langle m_{ee} \rangle$/[eV] & $0.00325$ \\ \hline
$\sum m_i$/[eV] & $0.0588$ \\ \hline \hline
$\chi^2$ & 16.9 \\  \hline
\end{tabular}
\caption{Benchmark point for model (II)}
\label{tab:3}
\end{table}
\end{center}
%
\begin{center} 
\begin{table}[tb]
\begin{tabular}{|c|c|}\hline 
$\tau$ & $-4.86 \times 10^{-5} + 1.05 \, i$ \\ \hline
$v_\chi/ [{\rm GeV}]$ & $1.37 \times 10^{14}$  \\ \hline
$[\sin \beta, \sin a, \sin b]$ & $[0.9910, 0.9980, 0.9710]$ \\ \hline
$[a_1, a_2, a_3]$ & $[-0.0817-0.0264 \, i, 0.281+0.164 \, i, 0.0103+0.0613 \, i]$ \\ \hline
$[a_4, a_5, a_6]$ & $[-0.00275-0.00116 \, i, -6.27-0.00116 \, i, 0.00436+0.00453 \, i]$ \\ \hline
$[b_1, b_2, b_3]$ & $[(4.11-7.02 \, i) \times 10^{-4}, -0.117-0.166 \, i, -0.00113-0.0173 \, i]$ \\ \hline
$[b_4, b_5, b_6]$ & $[(4.20+6.55 \, i) \times 10^{-4}, -0.471+0.0836 \, i, -0.00875+0.00127 \, i]$ \\ \hline
$[c_1, c_2, c_3]$ & $[0.250-0.000482 \, i, (6.21+4.28 \, i)\times 10^{-4}, -2.40 \times 10^{-7}+0.00719 \, i]$ \\ \hline
$[d_1, d_2, d_3]$ & $[-(12.5-2.13 \, i)\times 10^{-4}, (21.1+2.25 \, i)\times 10^{-5}, (5.77 - 401 \, i)\times 10^{-5}]$ \\ \hline
$[e_1, e_2, e_3]$ & $[-(4.24-21.5 \, i)\times 10^{-4}, (225+7.27 \, i) \times 10^{-4}, 0.643+0.0792 \, i]$ \\ \hline
$[e_4, e_5, e_6]$ & $[(2.24+75.7 \, i)\times 10^{-4}, -0.0231-1.29 \times 10^{-5} \, i, -0.182-7.34 \times 10^{-5} \, i]$ \\ \hline
$[\alpha_N, \beta_N]$ & $[-(7.64-431 \, i)\times 10^{-6}, -0.182-7.34 \times 10^{-4} \, i]$ \\ \hline \hline
$\bigl[ |V_{us}|,|V_{cb}|,|V_{ub}| \bigr]$ & $[0.224, 0.0424, 0.00356]$ \\ \hline
$\bigl[ |V_{td}|,|V_{ub}/V_{cb}| \bigr]$ & $[0.00850, 0.0840]$ \\ \hline
$\delta_{CP}^{CKM}/ [{\rm deg}]$ & 62.9 \\ \hline 
$[\sin^2 \theta_{12}, \sin^2 \theta_{23}, \sin^2 \theta_{13}]$ & $[0.304, 0.570, 0.0222]$ \\ \hline
$[\delta_{CP}, \alpha_{12}, \alpha_{13}]/ [{\rm deg}]$ & $[319, 118,140]$ \\ \hline 
$\langle m_{ee} \rangle$/[eV] & $0.00252$ \\ \hline
$\sum m_i$/[eV] & $0.0588$ \\ \hline \hline
$\chi^2$ & 4.49 \\ \hline
\end{tabular}
\caption{Benchmark point for model (III)}
\label{tab:4}
\end{table}
\end{center}
%
\begin{center} 
\begin{table}[tb]
\begin{tabular}{|c|c|}\hline 
$\tau$ & $3.14 \times 10^{-4} + 4.33 \, i$ \\ \hline
$v_\chi/ [{\rm GeV}]$ & $3.04 \times 10^{13}$ \\ \hline
$[\sin \beta, \sin a, \sin b]$ & $[0.999, 0.894, 0.448]$ \\ \hline
$[a_1, a_2, a_3]$ & $[0.00130+1.62 \, i, (8.62-202 \, i)\times 10^{-4}, 0.0547+0.00277 \, i]$ \\ \hline
$[a_4, a_5, a_6]$ & $[0.851-2.91 \times 10^{-5} \, i, -0.785+6.63 \times 10^{-5} \, i, 2.72 \times 10^{-4} + 1.21 \, i]$ \\ \hline
$[b_1, b_2, b_3]$ & $[9.08 \times 10^{-6}+0.342 \, i, (7.30-486 \, i)\times 10^{-4}, -0.124-0.00241 \, i]$ \\ \hline
$[b_4, b_5, b_6]$ & $[9.03 \times 10^{-6} - 30.5 \, i, 1.83-0.0501 \, i, 5.16 \times 10^{-6} - 3.63 \, i]$ \\ \hline
$[c_1, c_2, c_3]$ & $[-0.0115+0.326 \, i, -(3.36-3.33 \, i)\times 10^{-4}, 0.00185-0.00337 \, i]$ \\ \hline
$[c_4, c_5, c_6]$ & $[-9.33+2.95 \times 10^{-5} \, i, (7.21-829 \, i) \times 10^{-4}, 0.121+0.00329 \, i]$ \\ \hline
$[d_1, d_2, d_3]$ & $[-(6.02+125 \, i)\times 10^{-4}, -1.47 \times 10^{-6}+0.00121 \, i, -0.0539+0.00596 \, i]$ \\ \hline
$[d_4, d_5, d_6]$ & $[0.000964+8.81 \times 10^{-7} \, i, -(5.59-3.80 \, i)\times 10^{-6}, -7.79-2.01 \, i ]$ \\ \hline
$[e_1, e_2, e_3]$ & $[0.0829+0.0161 \, i, -0.00273-0.0115 \, i, (4.85+ 4.82 \, i)\times 10^{-2}]$ \\ \hline
$[e_4, e_5, e_6]$ & $[-0.0493-8.16 \, i, 4.22+1.52 \times 10^{-6}, -0.143+0.543 \, i]$ \\ \hline
$[\alpha_N, \beta_N]$ & $[0.00131-8.89 \times 10^{-7} \, i, 0.00717+0.00320 \, i]$ \\ \hline \hline
$\bigl[ |V_{us}|,|V_{cb}|,|V_{ub}| \bigr]$ & $[0.224, 0.0420, 0.00354]$ \\ \hline
$\bigl[ |V_{td}|,|V_{ub}/V_{cb}| \bigr]$ & $[0.00862, 0.0843]$ \\ \hline
$\delta_{CP}^{CKM}/ [{\rm deg}]$ & 66.0 \\ \hline 
$[\sin^2 \theta_{12}, \sin^2 \theta_{23}, \sin^2 \theta_{13}]$ & $[0.304, 0.570, 0.0222]$ \\ \hline
$[\delta_{CP}, \alpha_{12}, \alpha_{13}]/ [{\rm deg}]$ & $[344, 191,354]$ \\ \hline 
$\langle m_{ee} \rangle$/[eV] & $0.000423$ \\ \hline
$\sum m_i$/[eV] & $0.0606$ \\ \hline \hline
$\chi^2$ & 4.25 \\ \hline
\end{tabular}
\caption{Benchmark point for model (IV)}
\label{tab:5}
\end{table}
\end{center}
%

\section{Conclusion and discussion}
\label{sec:conclusion}

We have discussed models based on the $SU(4)_C \times SU(2)_L \times U(1)_R$ gauge symmetry and the $A_4$ modular flavor symmetry
that realize quark-lepton flavor unification.
The models are distinguished by the assignment of modular weight on superfields, which results in different structures of Yukawa couplings.
In these models, the active neutrino masses and the mixings are realized by the inverse seesaw mechanism, where the relevant mass matrix is determined by the modular symmetry.

We have carried out a numerical optimization of $\chi^2$ 
and checked which models can reproduce the observed quantities.
It is found that model (I) cannot accommodate the data of experimental measurements since it has less number of free parameters 
while we can fit the data in model (II), (III) and (IV).
Then we have shown a bestfit BP for each model that can accomodate all the masses and mixings in the fermion sector.
In model (II) and (III), we have found BPs with modulus $\tau$ close to the fixed point $\tau = i$ 
while we have obtained $\tau$ close to the fixed point $\tau = i \times \infty $ for the BP of model (IV).
Although there are many free parameters, model (IV) is an interesting possibility 
since it satisfies the criterion of scenarios in which the modular flavor symmetry is obtained from a higher-dimensional 
theory and the modular forms are understood as wave functions in extra dimensions.
It enables us to connect our model with higher dimensional scenarios such as stringy ones.

\section*{Acknowledgments}
\vspace{0.5cm}
{\it
This research was supported by an appointment to the JRG Program at the APCTP through the Science and Technology Promotion Fund and Lottery Fund of the Korean Government. This was also supported by the Korean Local Governments - Gyeongsangbuk-do Province and Pohang City (H.O.). 
H. O. is sincerely grateful for the KIAS member, and log cabin at POSTECH to provide nice space to come up with this project. OP is supported by the National Research Foundation of Korea Grants No. 2017K1A3A7A09016430 and No. 2017R1A2B4006338. 
The work was also supported by the Fundamental Research Funds for the Central Universities (T.~N.). Y.S.~is supported by I-CORE Program of the Israel Planning Budgeting Committee (grant No.\ 1937/12).
The authors gratefully acknowledge the computational and data resources provided by the Fritz Haber Center for Molecular Dynamics.
}

\appendix

\section{Formulas in modular $A_4$ framework}

In this appendix, we summarize relevant formulas in the $A_4$ modular symmetry framework. 
A modular form is a holomorphic function of modulus $\tau$, $f(\tau)$, that transforms as
\begin{align}
& \tau \rightarrow \gamma\tau= \frac{a\tau + b}{c \tau + d}\ ,~~ {\rm where}~~ a,b,c,d \in \mathbb{Z}~~ {\rm and }~~ ad-bc=1,  ~~ {\rm Im} [\tau]>0 ~, \\
& f(\gamma\tau)= (c\tau+d)^k f(\tau)~, ~~ \gamma \in \Gamma(N)~ ,
\end{align}
where $k$ is an integer called the modular weight and $\Gamma(N)$ ($N=3$ for $A_4$) is  an infinite normal subgroup of the group of $2\times2$ matrices with integer entries and determinant equal to one; $SL(2,Z)$.

Under the modular transformation, a superfield $\phi^{(I)}$ transforms such that
\begin{equation}
\phi^{(I)} \to (c\tau+d)^{-k_I}\rho^{(I)}(\gamma)\phi^{(I)},
\end{equation}
where  $-k_I$ is the modular weight of the superfield and $\rho^{(I)}(\gamma)$ is a representation matrix in the unitary representation of the $A_4$ transformation.
Therefore the superpotential is invariant if the sum of the modular weights of superfields and modular form is zero for each term (also it should be invariant under the $A_4$ symmetry and the gauge symmetry).

We can write the basis of modular forms with weight 2 as $ Y^{(2)}_3=  (y_{1},y_{2},y_{3})$ that transforms
as a triplet of $A_4$ and it is written in terms of the Dedekind eta-function $\eta(\tau)$ and its derivative \cite{Feruglio:2017spp}:
\begin{eqnarray} 
\label{eq:Y-A4}
y_{1}(\tau) &=& \frac{i}{2\pi}\left( \frac{\eta'(\tau/3)}{\eta(\tau/3)}  +\frac{\eta'((\tau +1)/3)}{\eta((\tau+1)/3)}  
+\frac{\eta'((\tau +2)/3)}{\eta((\tau+2)/3)} - \frac{27\eta'(3\tau)}{\eta(3\tau)}  \right), \nonumber \\
y_{2}(\tau) &=& \frac{-i}{\pi}\left( \frac{\eta'(\tau/3)}{\eta(\tau/3)}  +\omega^2\frac{\eta'((\tau +1)/3)}{\eta((\tau+1)/3)}  
+\omega \frac{\eta'((\tau +2)/3)}{\eta((\tau+2)/3)}  \right) , \label{eq:Yi} \\ 
y_{3}(\tau) &=& \frac{-i}{\pi}\left( \frac{\eta'(\tau/3)}{\eta(\tau/3)}  +\omega\frac{\eta'((\tau +1)/3)}{\eta((\tau+1)/3)}  
+\omega^2 \frac{\eta'((\tau +2)/3)}{\eta((\tau+2)/3)}  \right)\,, \nonumber \\
 \eta(\tau) &=& q^{1/24}\Pi_{n=1}^\infty (1-q^n), \quad q=e^{2\pi i \tau}, \quad \omega=e^{2\pi i /3}.
\nonumber
\end{eqnarray}
%
Modular forms with higher weight can be derived from the basis, $y_{1,2,3}(\tau)$, in use of the $A_4$ multiplication rules as shown below.
Thus, some $A_4$ triplet modular forms used in our analysis are derived as follows: 
\begin{align}
Y^{(4)}_3&\equiv (y^{(4)}_1,y^{(4)}_2,y^{(4)}_3) = ( y^2_1- y_2 y_3, y_3^2- y_1y_2,y_2^2- y_1y_3),\\
Y^{(6)}_3&\equiv (y^{(6)}_1,y^{(6)}_2,y^{(6)}_3) = ( y^3_1+2y_1 y_2 y_3, y_1^2y_2+2 y^2_2 y_3, y^2_1 y_3+2 y^2_3 y_2),\\
Y^{(6)}_{3'}&\equiv (y^{'(6)}_1,y^{'(6)}_2,y^{'(6)}_3) = ( y^3_3+2y_1 y_2 y_3, y^2_3 y_1+2 y^2_1 y_2, y^2_3 y_2+2 y^2_2 y_1).
\end{align}
The $A_4$ multiplication rules are summarized as  
\begin{align}
\begin{pmatrix}
a_1\\
a_2\\
a_3
\end{pmatrix}_{\bf 3}
\otimes 
\begin{pmatrix}
b_1\\
b_2\\
b_3
\end{pmatrix}_{\bf 3_2}
&=\left (a_1b_1+a_2b_3+a_3b_2\right )_{\bf 1} 
\oplus \left (a_3b_3+a_1b_2+a_2b_1\right )_{{\bf 1}'} \nonumber \\
& \oplus \left (a_2b_2+a_1b_3+a_3b_1\right )_{{\bf 1}''} \nonumber \\
&\oplus \frac13
\begin{pmatrix}
2a_1b_1-a_2b_3-a_3b_2 \\
2a_3b_3-a_1b_2-a_2b_1 \\
2a_2b_2-a_1b_3-a_3b_1
\end{pmatrix}_{{\bf 3}}
\oplus \frac12
\begin{pmatrix}
a_2b_3-a_3b_2 \\
a_1b_2-a_2b_1 \\
a_3b_1-a_1b_3
\end{pmatrix}_{{\bf 3_2}\  } \ , \nonumber \\
{(a)_{1'} } \otimes 
\begin{pmatrix}
b_1\\
b_2\\
b_3
\end{pmatrix}_{\bf 3}
&=a\begin{pmatrix}
b_3\\
b_1\\
b_2
\end{pmatrix}_{\bf 3},\quad
{(a)_{1''} } \otimes 
\begin{pmatrix}
b_1\\
b_2\\
b_3
\end{pmatrix}_{\bf 3}
=a\begin{pmatrix}
b_2\\
b_3\\
b_1
\end{pmatrix}_{\bf 3},
\nn\\
{\bf 1} \otimes {\bf 1} = {\bf 1} \ , \quad &
{\bf 1'} \otimes {\bf 1'} = {\bf 1''} \ , \quad
{\bf 1''} \otimes {\bf 1''} = {\bf 1'} \ , \quad
{\bf 1'} \otimes {\bf 1''} = {\bf 1} \ .
\end{align}

\end{document}